\documentclass[letterpaper,11pt]{article}
\usepackage[left=1in,right=1in,top=1in,bottom=1in]{geometry}
\usepackage[doublespacing]{setspace}
\usepackage{pdflscape}

\usepackage{soul,color}
\definecolor{seagreen}{RGB}{46, 139, 87}
\definecolor{coral}{RGB}{234,112,112}

\usepackage[reqno]{amsmath}
\usepackage[hidelinks]{hyperref}
\usepackage{amsthm}
\usepackage{amssymb,enumerate}
\usepackage{tikz}
\usetikzlibrary{fit,positioning,shapes,positioning,decorations.pathmorphing, decorations.pathreplacing,arrows}
\newcommand{\indep}{{\bot\negthickspace\negthickspace\bot}}
\newcommand{\E}{{\rm I\kern-.3em E}}

\renewcommand{\P}{\text{P}}

\usepackage{framed}
\usepackage{caption}
\usepackage[authoryear,sort]{natbib}
\title{\singlespacing Non-existent outcomes in research on inequality: A causal approach\thanks{An R package in development is available at \href{https://ilundberg.github.io/pstratreg}{\textcolor{blue}{ilundberg.github.io/pstratreg}}. For helpful discussions and feedback relevant to this project, we thank Brandon Stewart, Jennie Brand, Chris Felton, and members of the Inequality Data Science Lab at UCLA. This research benefited from feedback in presentations at the Cornell Center for the Study of Inequality, the UCLA Department of Sociology, and the American Causal Inference Conference. The authors benefited from facilities and resources provided by the California Center for Population Research at UCLA (CCPR), which receives core support (P2C-HD041022) from the Eunice Kennedy Shriver National Institute of Child Health and Human Development (NICHD). The content is solely the responsibility of the authors and does not necessarily represent the official views of the Eunice Kennedy Shriver National Institute of Child Health \& Human Development or the National Institutes of Health.}}

\author{Ian Lundberg\footnote{UCLA Department of Sociology and California Center for Population Research, \href{https://www.ianlundberg.org}{ianlundberg.org},  ianlundberg@ucla.edu.}\qquad Soonhong Cho\footnote{UCLA Department of Political Science, \href{https://soonhong-cho.github.io/}{soonhong-cho.github.io}, tnsehdtm@gmail.com.}}

\date{\today}

\begin{document}

\thispagestyle{empty}
\maketitle
\begin{center}

\end{center}

\begin{abstract}
\singlespacing
Scholars of social stratification often study exposures that shape life outcomes. But some outcomes (such as wage) only exist for some people (such as those who are employed). We show how a common practice---dropping cases with non-existent outcomes---can obscure causal effects when a treatment affects both outcome existence and outcome values. The effects of both beneficial and harmful treatments can be underestimated. Drawing on existing approaches for principal stratification, we show how to study (1) the average effect on whether an outcome exists and (2) the average effect on the outcome among the latent subgroup whose outcome would exist in either treatment condition. To extend our approach to the selection-on-observables settings common in applied research, we develop a framework involving regression and simulation to enable principal stratification estimates that adjust for measured confounders. We illustrate through an empirical example about the effects of parenthood on labor market outcomes.
\end{abstract}

\clearpage



\section{Introduction}
Many social and economic outcomes exist only for some people. Only employed individuals have wages, only married people can report marital satisfaction, and only those with children can transmit socioeconomic advantages to their descendants. Scholars studying these outcomes often restrict their analysis to those for whom the outcomes exist: the employed, the married, and the parents. Yet treatments that shape the outcome values often also affect whether those outcomes exist at all. 

We show how the standard practice of restricting analysis to those with observed outcomes can obscure causal effects, and we provide tools to resolve this problem. We focus on settings where a binary treatment shapes both (1) whether an outcome exists and (2) the value the outcome would take if it were to exist. Drawing on ideas for principal stratification developed for randomized experiments \citep{frangakis2002}, we define two quantities that researchers might want to study: the average causal effect on outcome existence and the average causal effect on outcome value among the latent subgroup who would have an outcome regardless of treatment condition. To adapt these methods to the selection-on-observables research designs common in quantitative sociology, we develop a new framework using regression to adjust for measured confounders and simulation to carry out principal stratification. We illustrate our approach with an application to the causal effect of motherhood on women's employment and hourly wages.

\section{Conceptual Framework}

Causal exposures often cause outcomes to come into existence or to cease to exist. This section illustrates the misleading conclusions that can arise when researchers focus solely on those whose outcomes exist.

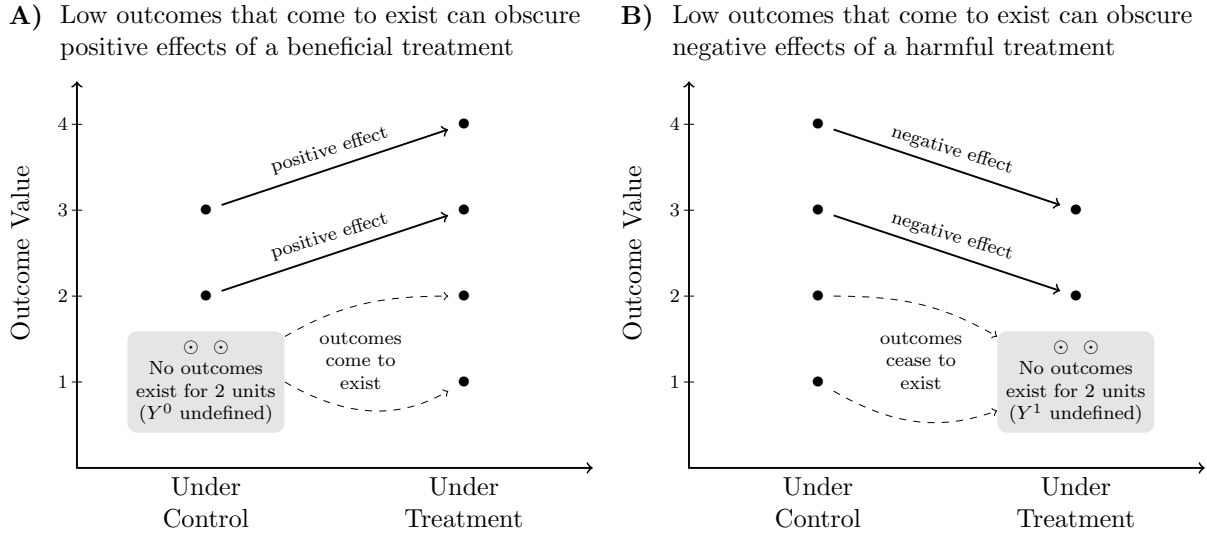
\begin{figure}
    \scalebox{.9}{\begin{tikzpicture}[x = 1.5in, y = .5in]
        \node[anchor = north west] at (-.8,5.5) {\textbf{A)}};
        \node[anchor = north west, align = left] at (-.6,5.5) {Low outcomes that come to exist can obscure\\positive effects of a beneficial treatment};
        \node at (0,4.8) {};
        \node at (0,-1) {};
        \node (y10) at (0,3) {$\bullet$};
        \node (y20) at (0,2) {$\bullet$};
        \node (y11) at (1,4) {$\bullet$};
        \node (y21) at (1,3) {$\bullet$};
        \node (y31) at (1,2) {$\bullet$};
        \node (y41) at (1,1) {$\bullet$};
        \draw[->, thick] (y10) -- node[midway, above, font = \scriptsize, rotate = 17] {positive effect} (y11);
        \draw[->, thick] (y20) -- node[midway, above, font = \scriptsize, rotate = 17] {positive effect} (y21);
        \node[anchor = north, align = center] at (0,0) {Under\\Control};
        \node[anchor = north, align = center] at (1,0) {Under\\Treatment};
        \draw[<->, thick] (-.5,4.5) -- (-.5,0) -- (1.5,0);
        \foreach \i in {1,2,3,4} {
            \node[anchor = east, font = \scriptsize] at (-.5,\i) {\i};
            \draw (-.52,\i) -- (-.48,\i);
        }
        \node[font = \scriptsize, align = center] at (.6,1.25) {outcomes\\come to\\exist};
        \node[anchor = south, rotate = 90] at (-.65,2.5) {Outcome Value};
        \node[fill = gray, fill opacity = .2, text opacity = 1, align = center, font = \scriptsize, rounded corners] (non) at (0,1) {$\odot$\hspace{6pt}$\odot$\\No outcomes\\exist for 2 units\\($Y^0$ undefined)};
        \draw[->, dashed] (non) to [out = 30, in = 180] (y31);
        \draw[->, dashed] (non.east) to [out = 330, in = 210] (y41);
    \end{tikzpicture}
    }
    \scalebox{.9}{\begin{tikzpicture}[x = 1.5in, y = .5in]
        \node[anchor = north west] at (-.8,5.5) {\textbf{B)}};
        \node[anchor = north west, align = left] at (-.6,5.5) {Low outcomes that come to exist can obscure\\negative effects of a harmful treatment};
        \node at (0,4.8) {};
        \node at (0,-1) {};
        \node (y11) at (1,3) {$\bullet$};
        \node (y21) at (1,2) {$\bullet$};
        \node (y10) at (0,4) {$\bullet$};
        \node (y20) at (0,3) {$\bullet$};
        \node (y30) at (0,2) {$\bullet$};
        \node (y40) at (0,1) {$\bullet$};
        \draw[->, thick] (y10) -- node[midway, above, font = \scriptsize, rotate = -17] {negative effect} (y11);
        \draw[->, thick] (y20) -- node[midway, above, font = \scriptsize, rotate = -17] {negative effect} (y21);
        \node[anchor = north, align = center] at (0,0) {Under\\Control};
        \node[anchor = north, align = center] at (1,0) {Under\\Treatment};
        \draw[<->, thick] (-.5,4.5) -- (-.5,0) -- (1.5,0);
        \foreach \i in {1,2,3,4} {
            \node[anchor = east, font = \scriptsize] at (-.5,\i) {\i};
            \draw (-.52,\i) -- (-.48,\i);
        }
        \node[anchor = south, rotate = 90] at (-.65,2.5) {Outcome Value};
        \node[fill = gray, fill opacity = .2, text opacity = 1, align = center, font = \scriptsize, rounded corners] (non) at (1,1) {$\odot$\hspace{6pt}$\odot$\\No outcomes\\exist for 2 units\\($Y^1$ undefined)};
        \node[font = \scriptsize, align = center] at (.4,1.25) {outcomes\\cease to\\exist};
        \draw[->, dashed] (y30) to [out = 0, in = 150]  (non);
        \draw[->, dashed] (y40) to [out = 330, in = 200] (non);
    \end{tikzpicture}
    }
    \caption{\textbf{Non-existent outcomes can obscure beneficial or harmful effects.} In Panel A, a treatment with positive effects also causes low outcomes to come to exist. A job training program lifts wages for two workers while helping two lower-paid workers to find jobs. In Panel B, a treatment with negative effects also causes low outcomes to cease to exist. Motherhood causes two women's wages to decline and causes two other women to leave paid employment entirely. In both panels, the mean value of the observed outcomes takes the same value (2.5) in both the control and the treatment conditions.}
    \label{fig:theory}
\end{figure}

Figure~\ref{fig:theory}A illustrates four hypothetical people who could receive treatment (job training). Two would have lower wages without job training and higher wages with job training. The other two would be non-employed without job training but would become employed with low wages with job training. A researcher who naively compared the average observed wage under treatment and control, however, would conclude that there was no effect: the observed mean wage among those employed is the same under treatment and control. Thus, a treatment that benefits everyone---either by improving wages or inducing employment---appears (misleadingly) to have zero average effect.

Figure~\ref{fig:theory}B illustrates four hypothetical employees facing potential firm downsizing. Two would have higher wages under no change but would experience wage declines under downsizing, while the other two would have lower wages under no change but would lose their jobs if the firm downsized. The naive comparison of average observed wages would again be misleading. A treatment (downsizing) that harms all four workers would appear to have zero average causal effect.

The hypothetical examples in Figure~\ref{fig:theory} are carefully designed to illustrate the point, but correspond to processes that are widespread in social stratification. As in Panel A, treatments that improve outcomes often bring new people into the population whose outcome values are low. In labor markets, job training programs raise wages for employed participants while helping previously unemployed people secure positions, typically at lower wages \citep{schochet2008does, card2018works}. Community investment programs in disadvantaged neighborhoods can raise wages for some residents while helping others avoid incarceration through new employment opportunities \citep{western2006}, and these otherwise-incarcerated people might join the pool of employed people at low wages. In educational contexts, tutoring programs boost achievement among enrolled students and prevent at-risk students from dropping out \citep{nickow2024promise}, while college completion improves earnings and increases employment rates \citep{brand2010benefits, brand2023overcoming}. In each case, the entry of these individuals with lower outcomes into the population of study can obscure the positive effects of the treatment.

As in Panel B, the negative impact of harmful treatments can also be obscured when these treatments cause low outcomes to disappear. In labor markets, automation may depress wages and also eliminate some of the lowest-wage positions held by lower-skilled workers \citep{autor2003skill, acemoglu2019automation}. Employment rates that exclude the incarcerated may underestimate the negative effects of economic shocks on labor market outcomes if those same shocks cause the most vulnerable to engage in criminal activity and become incarcerated \citep{western2002impact}. In families, financial strain might harm marital quality while pushing the most fragile marriages toward dissolution; average marital quality may seem unchanged simply because the worst-off marriages cease to exist \citep{conger1990linking}. In urban sociology, local gentrification could weaken social ties among residents, while also displacing those with the most tenuous connections so that they are no longer observed in the neighborhood at all \citep{desmond2016evicted, hwang2014divergent}.

These patterns---beneficial treatments creating low outcomes and harmful treatments eliminating them---illustrate the problem of non-existent outcomes in research on inequality. Analyses focusing solely on existing outcomes may understate both positive and negative treatment effects. Addressing this challenge requires methodological tools designed to study the existence or non-existence of outcomes as part of the causal process. In the next section, we formalize this problem using the potential outcomes framework  \citep{neyman1923application, rubin1974estimating} and provide a concrete example examining parenthood's effects on employment and wages.

\subsection{Potential outcomes formalize the selection problem}

We illustrate with a concrete example motivated by studies of how parenthood affects wages differently by gender. The uneven impact of parenthood on wages is a main source of gender pay inequality: fathers generally do not see a wage penalty or even earn more \citep{killewald2013}, while mothers often face significant wage decreases \citep{budig2001, waldfogel1997effect, staff2012explaining, gough2013review, england2016highly}. It is common in this literature to either condition on employment by focusing the analysis on employed individuals with wages, or to impute wages for the non-employed (often with simply zero or with predicted values from assumed models on missingness). Both approaches involve steps that might make researchers and readers uneasy. Conditioning on employment creates selection bias because employment is itself affected by the treatment, making employed and non-employed groups non-comparable. Wage imputation requires strong assumptions about outcomes that fundamentally do not exist: someone who refuses to tell you their wage has an unobserved wage, but someone who is not employed truly has no wage at all. These standard approaches obscure important aspects of inequality and produce unwarranted confidence in estimated effects on wage outcomes.

We illustrate the problem using four hypothetical women: Maya, Nancy, Mia, and Nia (Figure~\ref{fig:names}). For analytical clarity, we assume we know their potential wages under both motherhood and non-motherhood conditions---information unavailable in real data but useful for understanding the underlying logic. 

\begin{figure}
    \centering
    \scalebox{.85}{\begin{tikzpicture}[x = 1in, y = .2in]
    \node[align = center] (a1) at (1, .5) {if a mother};
    \node[align = center] (a0) at (2, .5) {if not};
    \node[align = center] (effect) at (3,.5) {effect};
    \node[font = \Large] at (1.6,.5) {${}-{}$};
    \node[font = \Large] at (2.5,.5) {${}={}$};
    \node[anchor = south west] at (a1.north west) {Employment};
    \draw[thick] (a1.south west) -- (a1.south east);
    \draw[thick] (a0.south west) -- (a0.south east);
    \draw[thick] (effect.south west) -- (effect.south east);
    \node[align = center] (a1) at (4, .5) {if a mother};
    \node[align = center] (a0) at (5, .5) {if not};
    \node[align = center] (effect) at (6,.5) {effect};
    \node[font = \Large] at (4.6,.5) {${}-{}$};
    \node[font = \Large] at (5.5,.5) {${}={}$};
    \node[anchor = south west] at (a1.north west) {Hourly wage};
    \draw[thick] (a1.south west) -- (a1.south east);
    \draw[thick] (a0.south west) -- (a0.south east);
    \draw[thick] (effect.south west) -- (effect.south east);
    \node[anchor = west] at (-1.2,-1) {Maya};
    \node[anchor = west] at (-1.2,-2) {Nancy};
    \node[anchor = west] at (-1.2,-3) {Mia};
    \node[anchor = west] at (-1.2,-4) {Nia};
    \node[anchor = west] at (-.7,-1) {is a mother};
    \node[anchor = west] at (-.7,-2) {is not a mother};
    \node[anchor = west] at (-.7,-3) {is a mother};
    \node[anchor = west] at (-.7,-4) {is not a mother};
    \node[draw, inner sep = 2pt] at (1,-1) {1};
    \node at (2,-1) {1};
    \node at (3,-1) {0};
    \node at (1,-2) {1};
    \node[draw, inner sep = 2pt] at (2,-2) {1};
    \node at (3,-2) {0};
    \node[draw, inner sep = 2pt] at (1,-3) {0};
    \node at (2,-3) {1};
    \node at (3,-3) {-1};
    \node at (1,-4) {0};
    \node[draw, inner sep = 2pt] at (2,-4) {1};
    \node at (3,-4) {-1};
    \node[draw, inner sep = 2pt] at (4,-1) {\$30};
    \node at (5,-1) {\$40};
    \node at (6,-1) {-\$10};
    \node at (4,-2) {\$30};
    \node[draw, inner sep = 2pt] at (5,-2) {\$40};
    \node at (6,-2) {-\$10};
    \node[draw, inner sep = 2pt] at (4,-3) {??};
    \node at (5,-3) {\$20};
    \node at (6,-3) {??};
    \node at (4,-4) {??};
    \node[draw] at (5,-4) {\$20};
    \node at (6,-4) {??};
    \end{tikzpicture}}
    \caption{\textbf{Conceptual illustration: Non-existent outcomes can obscure the motherhood wage penalty.} For a hypothetical set of four women, the figure depicts potential employment and wage outcomes that each individual would realize as a mother and as a non-mother. Boxes denote factual outcomes that would be observed in data. In this illustration, motherhood reduces employment by 50 percentage points on average. The motherhood wage penalty is -\$10 for Maya and Nancy, but it is an undefined quantity for Mia and Nia because their wage as a mother does not exist; they would not be employed in that condition. Despite beginning with a perfectly matched set of two mothers and two non-mothers, a researcher who dropped the non-employed (Mia) would induce selection bias such that there would incorrectly appear to be zero motherhood wage penalty on average.}
    \label{fig:names}
\end{figure}

Maya and Nancy are two women whose potential hourly wages are identical. Motherhood has no effect on their employment: they would remain employed regardless of motherhood. However, motherhood has a causal effect on their hourly wages: Each would earn \$30 as a mother but \$40 if not a mother. The only difference between Maya and Nancy is their realized treatment status: Maya is a mother while Nancy is not. Consequently, a researcher would only observe outcomes in their respective treatment conditions. If the researcher knew that Maya and Nancy had identical potential outcomes, the researcher could correctly estimate their \$10 motherhood wage penalty by matching them and subtracting Nancy's wage as a non-mother from Maya's wage as a mother.

Mia and Nia present another scenario. Neither would work for pay if they became mothers, but both would work at \$20 per hour if they remained childless. For Mia and Nia, motherhood would have a large negative effect on employment. The effect of motherhood on hourly wage, however, is an undefined quantity: the difference between a non-existent wage as a non-employed mother and a \$20 per hour wage as an employed non-mother. For Mia and Nia, employment status, rather than wage, constitutes the relevant labor market outcome.

Because the motherhood wage penalty is undefined for Mia and Nia, the average treatment effect is undefined for the population. The average effect of motherhood on hourly wages may not be the right causal question in this population. Instead, two causal questions emerge. (1) What is the average effect of motherhood on employment? In this population, motherhood reduces employment by 50\% because it has an effect of -1 in half of the population (Mia and Nia) and has an effect of 0 in the other half (Maya and Nancy). (2) Among those who would be employed regardless, what is the average effect of motherhood on hourly wage? In the subpopulation of Maya and Nancy, motherhood reduces the hourly wage by \$10 on average.

\subsection{How standard practice can mislead}

Standard practice produce misleading conclusions in this setting. The boxed values in Figure~\ref{fig:names} show the outcomes observable to researchers. When Mia is a mother, motherhood prevents her employment, resulting in no observable wage. A researcher following standard practice would exclude her from the analysis, resulting in a dataset with only one employed mother (Maya, \$30/hour) and two employed non-mothers (Nancy, \$40/hour; Nia, \$20/hour). While these women have different outcomes, we assume for this illustration that all four are identical on observed pre-treatment variables. If conducting matching, for instance, there would be no way to know whether to match Maya to Nancy or to Nia.

Comparing the average wage of employed mothers (\$30) with employed non-mothers ($\frac{\$40 + \$20}{2}$ = \$30) would erroneously suggest no motherhood wage penalty. This conclusion is particularly troubling given the true causal structure: motherhood harms all four women's labor market outcomes, either preventing employment (Mia and Nia) or reducing wages (Maya and Nancy). By excluding non-employed women, researchers inadvertently condition on a post-treatment variable (employment status), inducing post-treatment bias \citep{montgomery2018conditioning}. 

\subsection{Heckman selection models: A solution that does not work in our setting}

Our example is similar to the classic selection problem in labor market studies: people can report wages only if they are employed \citep{gronau1974wage, winship1992models, blau2017gender}. 
A popular solution is the Heckman selection model \citep{heckman1979sample}, which begins by defining an outcome that exists for all units. For a non-employed mother, we might define her outcome $Y$ to equal the wage she would be paid if counterfactually employed. Under this perspective, outcomes exist for all units but are observed only for the employed. We do not adopt this solution for two reasons. First, we find it philosophically difficult to define potential wages for those who are not employed: doing so changes the outcome of interest from realized wages to potential wages that would be realized under employment. Our approach instead takes the realized wage as the outcome.

The second reason concerns causal identification: the selection model approach requires an instrument that affects employment but is independent of the potential wages that would be realized if employed. With such an instrument and additional assumptions, it becomes possible to reweight the cases with observed outcomes to draw inference about all cases. But we would argue that such an instrument rarely exists for the kinds of questions that arise in social stratification. For example, it is difficult to imagine a variable that strongly predicts employment but is independent of the wage that would be realized if employed, thus satisfying what is sometimes called the exclusion restriction. Such a variable is hard to imagine because many causes of employment also shape wages. We instead prefer a solution that focuses on realized wages and allows the full causal process shaping employment to also shape wages.

\subsection{Notation}
\label{sec:notation}

Before presenting our preferred approach, we define notation that will be used throughout the paper. Let $S^a$ and $Y^a$ be potential outcome existence and outcome value under treatment value $a$. In the example above, outcome existence $S$ corresponds to employment and the outcome value $Y$ corresponds to wage. For units with $S^a=0$ the outcome $Y^a$ does not exist. The treatment values $a$ indicate parenthood or non-parenthood in our example. Denote $\mu(a,\vec{x}) = \E(Y\mid S = 1, A = a,\vec{X} = \vec{x})$ the mean observed outcome among those whose outcome exists with treatment value $a$ and confounders $\vec{x}$. Let $\mu^a(\vec{x}) = \E(Y^a\mid S^0=S^1=1,\vec{X} = \vec{x})$ be the mean potential outcome under treatment value $a$ among those with confounder vector $\vec{x}$ whose outcome would exist under either treatment. Denote $\tau(\vec{X}) = \mu^1(\vec{X}) - \mu^0(\vec{X})$ the difference in these mean functions, corresponding to a conditional average causal effect on outcome values. We use subscripts ``Lower'' and ``Upper'' to refer to lower and upper bounds, e.g. $\mu^0_\text{Lower}\leq \mu^0(\vec{X})\leq \mu^0_\text{Upper}$. Let $\pi(a,\vec{x}) = \P(S = 1\mid A = a, \vec{X} = \vec{x})$ be the probability of outcome existence given treatment value $a$ and confounder vector $\vec{x}$. Let $\pi_{S^0=S^1=1}(\vec{x}) = \P(S^0=S^1=1\mid \vec{X} = \vec{x})$ be the probability of having an outcome regardless of treatment, conditional on confounder vector $\vec{x}$, and $\pi_{S^0=S^1=1\mid S = 1}(a,\vec{x}) = \P(S^0=S^1=1\mid S = 1, A = a, \vec{X} = \vec{x})$ be the probability of outcome existence under either treatment, conditional on observed outcome existence under treatment value $a$.

\subsection{Principal stratification provides better causal estimands when some outcomes do not exist}

The challenge exemplified by Mia and Nia---where treatment affects whether an outcome exists at all---has been well-studied in biostatistics and epidemiology for medical trials where some participants do not survive to the end of the trial (``truncation by death''), possibly due to the intervention itself \citep{zhang2003, cheng2006bounds, ding2011identifiability}. The analytical structure remains analogous; just as death renders health outcomes non-existent, non-employment likewise renders wages non-existent. Economists have applied similar methods to job-training programs where training influences both employment probability and wages, creating non-random selection into the observed wage sample \citep{frumento2012evaluating, zhang2009likelihood, lee2009bound}. Political scientists have used principal stratification to study racial disparities in police use of force, where interactions with police only exist among those stopped by police, with race potentially affecting both the stopping decision and subsequent force \citep{knox2020administrative}.\footnote{More political science examples can be found in \cite{slough2023phantom}.} Our approach builds upon these applications but makes a distinct contribution by using parametric models to adjust for measured confounders.

The approach recognizes that units belong to latent groups termed \textit{principal strata}, defined by their potential survival outcomes $S^1$ under treatment and $S^0$ under control. A unit ``survives'' if it has a defined outcome $Y$ at the end of the study. In our example, survival corresponds to employment. The four women in our example fall into two principal strata: Maya and Nancy would be employed regardless of motherhood ($S^1 = 1, S^0 = 1$), while Mia and Nia would be employed only as non-mothers ($S^1 = 0, S^0 = 1$). With binary treatment and employment status, four principal strata exist: employed regardless of motherhood, employed only as non-mothers, employed only as mothers, and non-employed regardless of motherhood. Only the first two appear in Figure~\ref{fig:names}.

The first causal estimand of interest is the average causal effect on the existence of the outcome (i.e., survival), which in our example is the average difference in employment that would be realized as a mother ($S^1$) and employment that would be realized as a non-mother ($S^0$).
\begin{equation}
    \E\left(S^1 - S^0\right)
\end{equation}

When the effect on survival is large, many units have an outcome under only one treatment condition. For these units, the average causal effect on the outcome $Y$ is undefined since either $Y^0$ or $Y^1$ does not exist. The insight of principal stratification is that a meaningful contrast $Y^1-Y^0$ can only be well-defined for the principal stratum whose outcome exists regardless of treatment ($S^1=S^0=1$).\footnote{Principal stratification extends beyond addressing non-existent outcomes, providing a framework for handling various post-treatment complications. The method's most prominent application appears in studies of imperfect compliance, where instrumental variables estimate effects for the ``complier'' principal stratum---those units whose treatment status would align with their assignment under both treatment and control conditions \citep{angrist1996identification}.} Principal stratification targets the average causal effect on the outcome among the latent subgroup who would have an outcome regardless of treatment.\footnote{This local causal estimand is termed the ``Survivor Average Causal Effect'' (SACE) in biostatistics.}
\begin{equation}
\tau = \E\bigg(\underbrace{Y^1 - Y^0}_{\substack{\text{Effect of Treatment}\\\text{on Outcome Value}}}\mid \underbrace{S^1 = S^0 = 1}_{\substack{\text{Among Those Whose Outcome}\\\text{Exists Regardless of Treatment}}}\bigg) \label{eq:tau}
\end{equation}
In our example, this is the average effect on wage $Y$ among those who would be employed regardless of parenthood.

Estimating $\tau$ is difficult for two reasons. First, as with all causal estimands, at most one of $Y^1$ or $Y^0$ (potential outcomes) is observed for any unit. Second, only one of $S^1$ or $S^0$ (potential outcome existence) is observed. In concrete terms, for an employed non-mother like Nia, we observe $S_\text{Nia}^0 = 1$ but cannot observe her counterfactual employment status as a mother $S_\text{Nia}^1 = 0$. Data therefore cannot tell us if Nia belongs to our target subgroup (those employed regardless of motherhood) or to the subgroup that creates problems for the analysis (employed only if a non-mother). Because the estimand involves counterfactuals for both $Y$ and $S$, identification requires assumptions beyond those typically used for causal effects.

\section{Assumptions for principal stratification}

Principal stratum causal effects like Eq.~\ref{eq:tau} can be point-identified only under very strong assumptions, but can be set-identified within bounds under weaker, more credible assumptions \citep{miratrix2018bounding, knox2020administrative}. The stronger our assumptions, the tighter the bounds we are able to construct. This section introduces three assumptions one might make: conditional exchangeability, monotonicity, and mean dominance.

\subsection{Assumption: Conditional exchangeability}
\label{sec:conditional_exchangeability}

We first assume conditional exchangeability (no unmeasured confounding): potential outcome existence is independent of treatment given measured confounders $\vec{X}$.
\begin{equation}
    \{S^0,S^1\}\indep A\mid \vec{X} \quad \text{for }a\in\{0,1\} \qquad \text{(conditional exchangeability for outcome existence)}
\end{equation}
This assumption enables identification of the average causal effect on outcome existence by conditioning on $\vec{X}$, by an identification result standard in observational causal inference.
\begin{equation}
    \E(S^1-S^0) = \overbrace{\E_{\vec{X}}}^{\substack{\text{Population}\\\text{Mean}}}\overbrace{\bigg(\underbrace{\pi(1,\vec{X})}_{\substack{\text{Among Treated}\\\text{Given }\vec{X}}} - \underbrace{\pi(0,\vec{X})}_{\substack{\text{Among Untreated}\\\text{Given }\vec{X}}}\bigg)}^{\substack{\text{Difference in Outcome}\\\text{Existence Probabilities}}} \label{eq:s_identified}
\end{equation}

Because $Y^a$ is only defined when $S^a = 1$, conditional exchangeability for outcomes is more complex. At each treatment value $a = 0$ and $a = 1$, we assume that the potential outcome value $Y^a$ is independent of treatment among those whose outcome would exist under that treatment condition ($S^a=1$).
\begin{align}
    &Y^a\indep A \mid S^a=1, \vec{X} \quad \text{for }a\in\{0,1\} \qquad \text{(conditional exchangeability for outcome value)}
\end{align}

The assumptions of conditional exchangeability are analogous to their forms used in standard causal inference designs that rely on selection on observables. Section~\ref{sec:dag} returns to these assumptions to discuss their plausibility in our concrete example using a Directed Acyclic Graph.


\subsection{Assumption: Monotonicity}

Monotonicity assumes treatment has a one-directional effect on outcome existence, taking one of two forms: positive or negative monotonicity.\footnote{Monotonicity assumption has been widely applied in causal inference under selection \citep{lee2009bound} and instrumental variables approaches \citep{imbens1994identification}.} Under positive monotonicity, treatment may cause an outcome to exist but never causes it to cease existing:
\begin{equation}
    S_i^1 \geq S_i^0 \quad\text{for all }i\qquad \text{(positive monotonicity)}
\end{equation}
For example, job training may increase employment but never decrease it: anyone employed without training would also be employed with training. This aids identification because when we observe an untreated unit with an observed outcome (employed without training, $A_i = 0, S_i^0 = 1$), positive monotonicity implies this unit would have an outcome under either treatment condition ($S_i^0=S_i^1=1$).

Under negative monotonicity, treatment may cause an outcome to cease existing but never to come into existence:
\begin{equation}
    S_i^1 \leq S_i^0 \quad\text{for all }i\qquad \text{(negative monotonicity)}
\end{equation}
In our motherhood example, this means motherhood may reduce employment but never causes non-employed women to enter paid employment. When we observe a treated unit whose outcome exists (employed mother, $A_i = 1, S_i^1=1$), negative monotonicity implies she would also be employed if she counterfactually had no child ($S_i^0=1$), placing her in our target subgroup.

\subsection{Assumption: Mean dominance}

Mean dominance assumes a certain ordering of potential outcomes across principal strata. For example, women employed regardless of motherhood may have higher potential wages as non-motherhood, on average, than those employed only as non-mothers. Formally:
\begin{equation}
    \E(Y^0\mid S^0=S^1=1,\vec{X} = \vec{x}) \geq \E(Y^0\mid S^0=1,S^1=0,\vec{X} = \vec{x})\quad\forall \vec{x} \qquad\text{(mean dominance)}
\end{equation}
Mean dominance can be positive or negative, for $Y^0$ or $Y^1$, depending on the setting. Here we focus on positive mean dominance for $Y^0$. This holds in our four-woman example, where always-employed women (Maya and Nancy, \$40) have higher potential non-mother hourly wages than those employed only as non-mothers (Mia and Nia, \$20).

Mean dominance would be plausible if women employed regardless of motherhood have characteristics associated with higher wages compared to those who would exit upon motherhood: better human capital (e.g. education, skills, work experience), stronger labor market attachment (e.g., career ambition, occupational choice), or more favorable job characteristics (e.g., family-friendly policies, flexible hours). The assumption could fail if low-wage women tend to continue working due to financial constraints, or if high-wage women exit employment upon motherhood---either because their jobs are more family-unfriendly or because they have higher-earning spouses providing financial flexibility. Researchers who make the assumption of mean dominance should present an argument for why it may hold in their particular setting.

\section{Bounding effects on $Y$: Nonparametric set identification}

Conditional exchangeability, monotonicity, and mean dominance enable nonparametric set identification of average causal effects among those who would have an outcome regardless of treatment. We first discuss a particularly tractable setting: bounding the average treatment effect on the treated under negative monotonicity. Then we discuss more general bounding results for average treatment effects. This section focuses on nonparametric identification, and the following section introduces parametric estimation.

\subsection{Average treatment effect on the treated under conditional exchangeability and negative monotonicity}

We first consider the average treatment effect on treated units who would have an outcome regardless of treatment (ATT). In our example, this is the effect of motherhood on wages among mothers who would be employed regardless of motherhood.
\begin{align}
\tau^\text{ATT} &= \E\bigg(\overbrace{Y^1 - Y^0}^{\substack{\text{Effect of Treatment}\\\text{on Outcome Value}\\\text{(e.g., wage)}}}\mid \overbrace{A = 1}^{\substack{\text{Among}\\\text{Treated Units}\\\text{(e.g., parents)}}}, \quad \overbrace{S^1 = S^0 = 1}^{\substack{\text{Whose Outcome Would Exist}\\\text{Regardless of Treatment}\\\text{(e.g., employed regardless)}}}\bigg) \\
&= \E_{\vec{X}}\left(\mu^1(\vec{X})\mid A = 1\right) - \E_{\vec{X}}\left(\mu^0(\vec{X})\mid A = 1\right) \label{eq:tau_att}
\end{align}
where the second line decomposes $\tau_\text{ATT}$ into two components using the law of iterated expectation and linearity of expectation.

The first component is the outcome under treatment for treated units whose outcome would exist regardless of treatment. Under negative monotonicity ($S_i^1\leq S_i^0$), any treated unit with an outcome ($S_i^1=1$) would also have an outcome in the absence of treatment ($S_i^0=1$). In our concrete example, Maya, an employed mother, would (by monotonicity) also be employed as a non-mother. Thus, the first component is point-identified as the mean among treated units with outcomes.
\begin{align}
    \E_{\vec{X}}\left(\mu^1(\vec{X})\mid A = 1\right) &= \E(Y^1\mid S^0=S^1=1,A=1) &\text{by iterated expectation}\\
    &= \E(Y^1\mid S =1 , A = 1) &\text{by negative monotonicity} \\
    &= \E(Y\mid S =1 , A = 1) &\text{by consistency}
\end{align}

The second component of $\tau^\text{ATT}$ is more challenging. Under conditional exchangeability, counterfactual outcomes (without motherhood) can be identified from comparable non-mothers (conditional on $\vec{X}$ and $S^0=S^1=1$).
\begin{align}
    &\E_{\vec{X}}\left(\mu^0(\vec{X})\mid A = 1\right) \nonumber \\     
    &=\E_{\vec{X}}\left(\E(Y^0\mid S^0=S^1=1,A=1, \vec{X})\right) &\text{by iterated expectation}\\
    &= \E_{\vec{X}}\left(\E(Y^0\mid S^0=S^1=1 , A = 0,\vec{X}) \right) &\text{by conditional exchangeability} \\
    &= \E_{\vec{X}}\bigg(\E(Y\mid \underbrace{S^0=S^1=1}_{\substack{\text{Difficulty:}\\\text{Latent Subgroup}}} , A = 0,\vec{X})\bigg) &\text{by consistency}
\end{align}

However, the conditioning on $S^0=S^1=1$ involves unobservable potential outcomes: for non-mothers, we cannot observe $S^1$. In our concrete example, we know Nancy ($S_\text{Nancy}^1=1$) belongs to our target stratum while Nia ($S_\text{Nia}^1=0$) does not, but this is unobservable in real data because $S^1$ is counterfactual for them. The challenge is that the employed non-mothers are a mixture of two principal strata: one stratum employed regardless (Nancy) and another stratum who would only be employed as non-mothers (Nia).

Under negative monotonicity, we can point-identify the proportion of employed non-mothers who are in the target stratum, within any subgroup taking a particular confounder vector value $\vec{X}$.
\begin{align}
    \overbrace{\pi_{S^0=S^1=1\mid S=1}(A = 0,\vec{X})}^{\substack{\text{Proportion Employed Regardless}\\\text{Among Employed Non-Mothers}}}
    &= \P(S^1=1\mid \vec{X}, A = 0, S^0 = 1) \\
    &= \frac{\P(S^1=1,S^0=1\mid \vec{X}, A = 0)}{\P(S^0=1\mid \vec{X}, A = 0)} &\text{by def. of conditional prob.} \\
    &= \frac{\P(S^1=1\mid \vec{X}, A = 0)}{\P(S^0=1\mid \vec{X}, A = 0)} &\text{by negative monotonicity} \\
    &= \frac{\P(S^1=1\mid \vec{X}, A = 1)}{\P(S^0=1\mid \vec{X}, A = 0)} &\text{by conditional exchangeability} \\
    &= \frac{\P(S=1\mid \vec{X}, A = 1)}{\P(S=1\mid \vec{X}, A = 0)} &\text{by consistency} \\
    &= \frac{\pi(1,\vec{X})}{\pi(0,\vec{X})}
\end{align}
Intuitively, under negative monotonicity, this equation says that the proportion who would have an outcome under $A = 1$, among those whose outcome exists under $A = 0$, equals the ratio of outcome existence under $A = 1$ to outcome existence under $A = 0$.

The second step is to use this proportion to bound the target estimand. Among employed non-mothers with covariate $\vec{X}$, a proportion $\pi(\vec{X})$ are in the stratum of interest and a proportion $1 - \pi(\vec{X})$ are not. We create lower (upper) bounds on $\mu^0(\vec{x})$ for all $\vec{x}$ by considering the extreme cases where the $\pi(\vec{X})$ fraction of units in the always-employed stratum are from the lowest- (highest-)valued portions of the observed distribution:
\begin{align}
    \mu^0_\text{Lower}(\vec{X}) &= \E\bigg(
        Y\mid S = 1, A = 0, \vec{X}, 
        \underbrace{F_{Y\mid S = 1,A = 0,\vec{X}}(Y) < 
     \pi_{S^1=1\mid S=1, A = 0}(\vec{X})}_{\substack{\text{averaging over the employed-regardless}\\\text{assumed to be the lower}\\\text{portion of the distribution}}}
    \bigg) \\
    \mu^0_\text{Upper}(\vec{X}) &= \E\bigg(
        Y\mid S = 1, A = 0, \vec{X}, 
        \underbrace{F_{Y\mid S = 1,A = 0,\vec{X}}(Y) > 
     1 - \pi_{S^1=1\mid S=1, A = 0}(\vec{X})}_{\substack{\text{averaging over the employed-regardless}\\\text{assumed to be the upper}\\\text{portion of the distribution}}}
    \bigg)
\end{align}
Finally, we bound $\tau_\text{ATT}$ by taking the difference between (1) the sample mean of the treated units' outcomes and (2) the bounds on the estimates for their outcome in the absence of treatment.

\begin{align}
    \E\bigg(Y - \mu^0_\text{Upper}(\vec{X})\mid S = 1, A = 1\bigg) \leq \tau^\text{ATT} &\leq \E\bigg(Y - \mu^0_\text{Lower}(\vec{X})\mid S = 1, A = 1\bigg) \label{eq:bounds_att}
\end{align}

Our four-women example illustrates these bounds. The mean outcome of employed mothers (Maya) is \$30. The employed non-mothers Nia (\$20) and Nancy (\$40) are a mixture: one is employed-regardless and one would leave employment if she became a mother. Without knowing which non-mother belongs to which group, we construct bounds by considering both possibilities: if Nancy is the always-employed non-mother, the effect is (Maya - Nancy) = (\$30 - \$40) = \$10; if Nia is the always-employed non-mother, the effect is (Maya - Nia) = (\$30 - \$20) = +\$10. Our assumptions and evidence thus bound the average effect between $-\$10$ and $+\$10$.

Bounding the ATT under negative monotonicity is straightforward because treated units with outcomes $\{i:S_i = 1, A_i = 1\}$ coincide exactly with the target population $\{i:S_i^0=S_i^1=1, A_i = 1\}$. This simplifies two steps. First, the expected outcome under treatment equals the observed mean of $Y$ among these units. Second, the conditional bounds $\mu^0_\text{Lower}(\vec{X})$ and $\mu^1_\text{Upper}(\vec{X})$ could be aggregated over the observed distribution of $\vec{X}$: the distribution $\vec{X}\mid S = 1, A = 1$ among treated units with existing outcomes (Eq.~\ref{eq:bounds_att}).

\subsection{Average treatment effect under negative monotonicity}

For the average treatment effect over all units who have outcomes regardless, two additional challenges arise: (1) estimating potential outcomes under treatment for control units, and (2) aggregating over the conditional distribution $\vec{X}\mid S^0=S^1=1$.

Under conditional exchangeability and negative monotonicity, the conditional mean outcome under treatment for those whose outcome exists regardless is point-identified.
\begin{align}
    &\E(Y^1\mid S^0=S^1=1, \vec{X}) \\
    &= \E(Y^1\mid S^0=S^1=1, A = 1, \vec{X}) &\text{by conditional exchangeability} \\
    &= \E(Y^\mid S^1=1, A = 1, \vec{X}) &\text{by negative monotonicity} \\
    &= \E(Y\mid S=1, A = 1, \vec{X}) &\text{by consistency} \\
    &= \mu^1(\vec{X})
\end{align}

Similarly, the conditional proportion whose outcome exists regardless given confounders $\vec{X}$ is identified by the conditional proportion of treated units who have an outcome.
\begin{align}
    \pi_{S^0=S^1=1}(\vec{X}) &= \P(S^0=S^1=1\mid\vec{X}) &\text{defining abbreviated notation} \\
    &= \P(S^1=1\mid\vec{X}) &\text{by negative monotonicity} \\
    &= \P(S^1=1\mid A = 1, \vec{X}) &\text{by conditional exchangeability}\\
    &= \P(S=1\mid A = 1, \vec{X}) &\text{by consistency}
\end{align}

These results bounds $\tau_\text{ATE}$ by the weighted average of conditional estimates, with weights proportional to stratum membership probabilities.
\begin{equation}
    \underbrace{\E_{\vec{X}}}_{\substack{\text{Mean}\\\text{over}\\\vec{X}}}\bigg(
    \underbrace{\tau_\text{Lower}(\vec{X})}_{\substack{\text{Lower bound}\\\text{on average}\\\text{effect given }\vec{X}}}
    \hspace{6pt}
    \underbrace{\frac{\pi_{S^0=S^1=1}(\vec{X})}{\text{E}(\pi_{S^0=S^1=1}(\vec{X}))}}_{\substack{\text{Weighted by the rate}\\\text{of having an outcome}\\\text{regardless of treatment}\\\text{given }\vec{X}}}\bigg) 
    \leq \tau_\text{ATE} \leq 
    \E\bigg(
    \underbrace{\tau_\text{Upper}(\vec{X})}_{\substack{\text{Upper bound}\\\text{on average}\\\text{effect given }\vec{X}}}
    \hspace{6pt}
    \underbrace{\frac{\pi_{S^0=S^1=1}(\vec{X})}{\text{E}(\pi_{S^0=S^1=1}(\vec{X}))}}_{\substack{\text{Weighted by the rate}\\\text{of having an outcome}\\\text{regardless of treatment}\\\text{given }\vec{X}}}
    \bigg)
\end{equation}

\subsection{Average treatment effect without assuming monotonicity}
\label{sec:ate_no_monotonicity}

Without monotonicity, set identification becomes more challenging because even the weight $\pi_{S^0=S^1=1}(\vec{X})$ is only set-identified:
\begin{equation}
    \pi_{\text{Lower},S^0=S^1=1}(\vec{x}) \quad \leq \quad \pi_{S^0=S^1=1}(\vec{x}) \quad \leq \quad \pi_{\text{Upper},S^0=S^1=1}(\vec{x}),
\end{equation}
with formulas for these bounds provided in Appendix \ref{sec:proof_no_monotonicity}. A general procedure is to first create bounds on the conditional average treatment effect and the proportion whose outcomes exist regardless, both conditional on $\vec{X}$:
\begin{align}
    \tau_\text{Lower}(\vec{X})
        \quad \leq \quad \E(Y^1-Y^0\mid S^0=S^1=1,\vec{X})
        \quad \leq \quad \tau_\text{Upper}(\vec{X}).
\end{align}

Aggregating these conditional bounds over $\vec{X}$ is challenging because the weight for each covariate value $\vec{x}$ depends on the proportion of units who would have an outcome regardless in that subgroup---a quantity that is only set-identified. We propose a sequential procedure that strategically allocates weights: lower bounds emerge by weighting observations with smaller conditional effects, while upper bounds prioritize observations with larger conditional effects.

For the lower bound: (1) Sort observations by $\tau_\text{Lower}(\vec{x}_i)$ from smallest to largest; (2) Initialize all stratum probabilities to $\tilde\pi(\vec{x}_i)=\pi_\text{Lower}(\vec{x}_i)$ and calculate the (tentative) initial weighted estimate $\tilde\tau_\text{Lower} = \frac{1}{\sum_i\tilde\pi(\vec{x}_i)}\sum_i \tilde\pi(\vec{x}_i)\tau_\text{Lower}(\vec{x}_i)$; (3) Starting with the smallest effect observation, sequentially increase each probability to its maximum, $\pi_\text{Upper}(\vec{x}_i)$, and re-calculate $\tilde\tau_\text{Lower}$ after each update; (4) Continue this process for observations with $\tau_\text{Lower}(\vec{x}_i)$ less than the current estimate until increasing weights would no longer reduce $\tilde\tau_\text{Lower}$. This yields a valid lower bound $\tau_\text{Lower} = \tilde\tau_\text{Lower}$.

For the upper bound, we carry out the same procedure in reverse, starting with observations having the largest $\tau_\text{Upper}(\vec{x}_i)$ values and sequentially increasing their weights to maximize the estimate.
While valid, we anticipate that its usefulness may be rare in practice as it may yield very wide bounds in many settings. Since monotonicity is often credible in social science applications and produces narrower, simpler bounds, we focus primarily on monotonicity-based bounds.

\section{Parametric estimation by regression and simulation}
\label{sec:parametric}

\begin{figure}[!htbp]
    \centering
    \resizebox{!}{.85\textheight}{
    \begin{tikzpicture}[x = \textwidth, y = .8\textheight]
        \node[anchor = north west] at (0,1) {\textbf{1)} Define causal estimands.};
        \node[anchor = north west, font = \footnotesize] at (.07,.96) {$\E(S^1-S^0)$};
        \node[anchor = north east, font = \footnotesize] at (1,.96) {(effect on outcome existence)};
        \node[anchor = north west, font = \footnotesize] at (.07,.93) {$\E(Y^1-Y^0\mid S^0=S^1=1)$};
        \node[anchor = north east, font = \footnotesize] at (1,.93) {(effect on $Y$ among those with outcomes regardless)};
        \node[anchor = north west] at (0,.89) {\textbf{2)} Make causal assumptions.};
        \node[anchor = west, font = \footnotesize] at (.07,.83) {$S^a\indep A\mid \vec{X}$};
        \node[anchor = west, font = \footnotesize] at (.3,.83) {for all $a$};
        \node[anchor = east, font = \footnotesize] at (1,.83) {(conditional exchangeability for survival)};
        \node[anchor = west, font = \footnotesize] at (.07,.8) {$Y^a\indep A\mid S^a=1, \vec{X}$};
        \node[anchor = west, font = \footnotesize] at (.3,.8) {for all $a$};
        \node[anchor = east, font = \footnotesize] at (1,.8) {(conditional exchangeability for outcome)};
        \node[anchor = west, font = \footnotesize] at (.07,.77) {$S_i^1 \geq S_i^0$};
        \node[anchor = west, font = \footnotesize] at (.3,.77) {for all $i$};
        \node[anchor = east, font = \footnotesize] at (1,.77) {(positive monotonicity)};
        \node[anchor = west] at (0,.73) {\textbf{3)} Make statistical modeling assumptions.};
        \node[anchor = north west, align = left, font = \footnotesize] at (.07,.7) {Model outcome existence:};
        \node[anchor = north west, align = left, font = \footnotesize] at (.07,.67) {$S\mid A,\vec{X}\sim \text{Bernoulli}\big(\pi(A,\vec{X})\big)$};
        \node[anchor = north west, align = left, font = \footnotesize] at (.5,.7) {Model outcome given its existence:};
        \node[anchor = north west, align = left, font = \footnotesize] at (.5,.67) {$Y \mid S = 1, A,\vec{X} \sim \text{Normal}\big(\mu(A,\vec{X}), \sigma^2(A,\vec{X})\big)$};
        \node[anchor = north west] at (.07, .63) {\includegraphics[width = .4\textwidth]{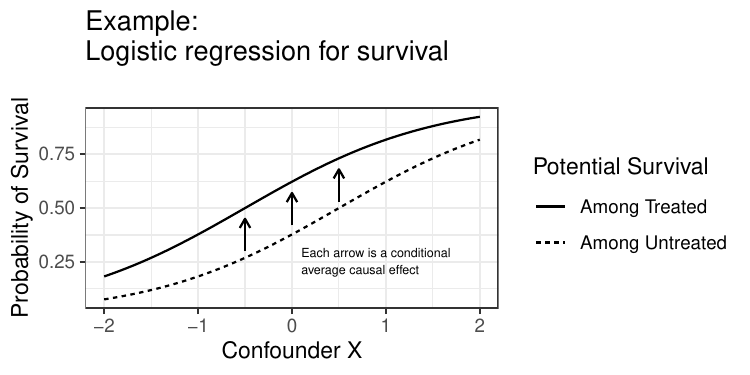}};
        \node[anchor = north west] at (.5, .63) {\includegraphics[width = .4\textwidth]{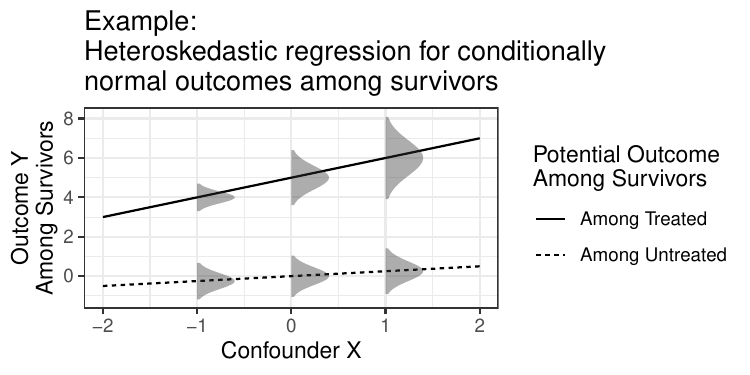}};
        \node[anchor = west] at (0,.41) {\textbf{4)} Simulate outcomes at each $\vec{X} = \vec{x}_i$. Bound $\mu^a(\vec{x}_i) = \E(Y^a\mid S^0=S^1=1,\vec{X} = \vec{x}_i)$.};
        \node[anchor = north west, align = left, font = \footnotesize] at (.07,.38) {Simulated $Y$ of \textbf{treated} survivors\\who observationally match unit $i$:};
        \node[anchor = north] at (.3,.34) {\includegraphics[width = .2\textwidth]{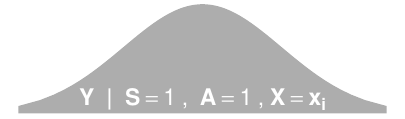}};
        \node[anchor = north west, align = left, font = \footnotesize] at (.53,.38) {Simulated $Y$ of \textbf{untreated} survivors\\who observationally match unit $i$:};
        \node[anchor = north] at (.75,.34) {\includegraphics[width = .2\textwidth]{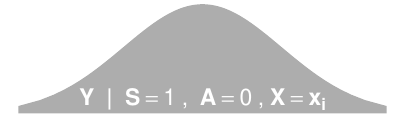}};
        \node[anchor = north west, align = left, font = \footnotesize] at (.07,.28) {By positive monotonicity, $\frac{\pi(0,\vec{x}_i)}{\pi(1,\vec{x}_i)}$ of this density\\are those who survive regardless};
        \node[anchor = north west, align = left, font = \footnotesize] at (.53,.28) {By positive monotonicity, 100\% of this density\phantom{$\frac{\pi(0,\vec{x}_i)}{\pi(1,\vec{x}_i)}$}\\are those who survive regardless};
        \node[anchor = north] (upper) at (.4,.22) {\includegraphics[width = .2\textwidth]{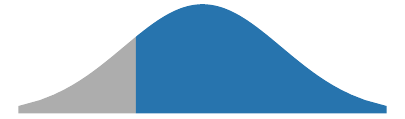}};
        \node[anchor = north] (lower) at (.2,.22) {\includegraphics[width = .2\textwidth]{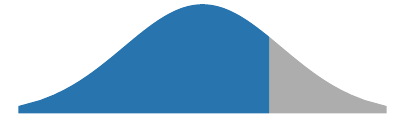}};
        \node[anchor = north, align = center, font = \footnotesize] at (upper.south) {Upper Bound:\\Mean of Blue Region\\$\mu^1_\text{Upper}(\vec{x}_i)$};
        \node[anchor = north, align = center, font = \footnotesize] at (lower.south) {Lower Bound:\\Mean of Blue Region\\$\mu^1_\text{Lower}(\vec{x}_i)$};
        \node[white, font = \tiny, shift={(.5ex,-.2ex)}] at (upper) {$\frac{\pi(0,\vec{x}_i)}{\pi(1,\vec{x}_i)}$};
        \node[white, font = \tiny, shift={(-.5ex,-.2ex)}] at (lower) {$\frac{\pi(0,\vec{x}_i)}{\pi(1,\vec{x}_i)}$};
        \node[anchor = north] (y0_all) at (.75,.22) {\includegraphics[width = .2\textwidth]{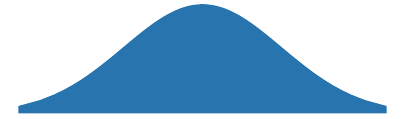}};
        \node[anchor = north, align = center, font = \footnotesize] at (y0_all.south) {Point Identified:\\Mean of Blue Region\\$\mu^0(\vec{x}_i)$};
        \draw[rounded corners] (.06,.07) rectangle (.52,.38);
        \draw[rounded corners] (.52,.07) rectangle (.98,.38);
        \node[anchor = north west] at (0,.05) {\textbf{5)} Difference to bound the conditional average effect on $Y$ among those who survive regardless.};
        \node[anchor = north, align = left, font = \footnotesize] at (.52,0.01) {$\hat\tau_\text{Lower}(\vec{x}_i) = \hat\mu^1_\text{Lower}(\vec{x}_i) - \hat\mu(0,\vec{x}_i) \qquad \qquad \hat\tau_\text{Upper}(\vec{x}_i) = \hat\mu^1_\text{Upper}(\vec{x}_i) - \hat\mu(0,\vec{x}_i)$};
        \node[anchor = west] at (0,-.05) {\textbf{6)} Aggregate over $\vec{X}$ to produce average causal effect estimates.};
        \node[anchor = west, font = \footnotesize] at (.07,-.09) {$\hat\E(S^1-S^0) \quad = \quad \frac{1}{n}\sum_{i=1}^n\left(\qquad \hat\pi(1,\vec{x}_i) \quad - \quad \hat\pi(0,\vec{x}_i)\qquad\right)$};
        \node[anchor = north, font = {\footnotesize}, scale = .8, align = center] at (.13,-.11) {effect on\\survival};
        \node[anchor = north, font = \footnotesize, scale = .8, align = center] at (.27,-.11) {average\\over units};
        \node[anchor = north, font = \footnotesize, scale = .8, align = center] at (.38,-.11) {predicted survival\\under treatment};
        \node[anchor = north, font = \footnotesize, scale = .8, align = center] at (.53,-.11) {predicted survival\\under control};
        \node[font = \footnotesize, anchor = west] at (.07,-.2) {$\hat\E(Y^1-Y^0\mid S^0=S^1=1) \in\bigg[\frac{1}{\sum_i \hat\pi(0,\vec{x}_i)} \sum_i \hat\pi(0,\vec{x}_i) \hat\tau^\text{Lower}(\vec{x}_i), \quad \frac{1}{\sum_i \hat\pi(0,\vec{x}_i)} \sum_i \hat\pi(0,\vec{x}_i)\hat\tau^\text{Upper}(\vec{x}_i)\bigg]$};
        \node[anchor = north, font = {\footnotesize}, scale = .8, align = center] at (.2,-.23) {effect on $Y$ among those\\whose outcome exists regardless};
        \node[anchor = north, font = {\footnotesize}, scale = .8, align = center] at (.61,-.23) {conditional bounds weighted by conditional rate of outcome existing regardless\\(under positive monotonicity)};
        \draw[rounded corners] (-0.03,-.28) rectangle (1.03,1.01);
    \end{tikzpicture}
    }
    \caption{\textbf{Principal stratification with regression: Illustrated with positive monotonicity.} In the figure, outcome $Y_i$ exists for any unit that survives ($S_i = 1)$, such as a wage existing only for employed individuals. Positive monotonicity assumes the intervention (e.g., $A_i = 1$ indicates receipt of job training) never decreases employment probability. Note that this figure illustrates positive monotonicity ($S^1 \geq S^0$) as applies in job training scenarios, whereas our motherhood application uses negative monotonicity ($S^0 \geq S^1$). Steps 4--6 become more complicated without monotonicity, because the conditional probability of surviving regardless is only set-identified.}
    \label{fig:overview_of_approach}
\end{figure}

In randomized settings or settings with few discrete confounders $\vec{X}$, the identification results can be applied with nonparametric plug-in estimators, as in \citep{zhang2003, miratrix2018bounding}. However, in many observational studies, confounders $\vec{X}$ may be continuous or high-dimensional. It may be unusual to observe multiple units taking a single value of the confounders, so that nonparametric estimation is simply infeasible. This section presents our main methodological contribution: an estimation strategy that relies on parametric regression and simulation (Figure~\ref{fig:overview_of_approach}).

\subsection{Parametrically model the existence of outcomes}

We model the probability of outcome existence using logistic regression with treatment-specific coefficients:
\begin{align}
    \pi(a, \vec{x}) = \text{logit}^{-1}\left(\alpha_a + \vec{x}'\vec\beta_a\right)
\end{align}

\subsection{{Parametrically model outcome values given existence}}

For outcome $Y$ among those for whom this outcome exists, we assume conditional normality with treatment-specific mean and variance functions. We estimate the variance model by a generalized linear model assuming a Gamma distribution on squared residuals \citep{western2009}.
\begin{align}
    Y\mid S = 1, A = a, \vec{X} = \vec{x} &\sim \mathcal{N}\bigg(\mu(a,\vec{x}), \sigma^2(a,\vec{x})\bigg) \\
    \mu(a,\vec{x}) &= \eta_a + \vec{X}'\vec\lambda_a \\
    \text{log}\big(\sigma^2(a,\vec{x})\big) &= \nu_a + \vec{X}'\vec\gamma_a
\end{align}

Two important considerations are relevant to this model. First, units with existing outcomes ($S = 1$) represent a mixture of latent principal strata in proportions that differ by treatment group. The simulation step that follows, thus, is essential to yield interpretable causal estimates from the outcome model. Second, while many researchers focus solely on conditional mean functions $\mu(a,\vec{x})$, our estimators require simulating from the full conditional outcome distribution. Hence, in this model the variance term $\sigma^2(a,\vec{x})$ and the assumption of conditional normality are also modeling assumptions of critical importance.

\subsection{Simulate to estimate principal stratification estimands}

The final step uses the estimated model to simulate principal stratification estimands. The first estimand of interest---the average causal effect on outcome existence---can be point estimated by predicting $\hat\pi(a,\vec{x}_i)$ from the model for outcome existence under each treatment value $a$, plugging these into Eq.~\ref{eq:s_identified}, and estimating the population mean with the sample mean.
\begin{align}
    \hat\E\left(S^1-S^0\right) &= \frac{1}{n}\sum_{i=1}^n \bigg(\hat\pi(1,\vec{x}_i) - \hat\pi(0,\vec{x}_i)\bigg)
\end{align}

Turning attention to outcomes, the simulation procedure varies by estimand and the assumptions maintained. For concreteness, here we consider ATT among those whose outcome would exist regardless, under negative monotonicity.
\begin{equation}
    \tau^\text{ATT} = \E(Y^1-Y^0\mid A = 1, S^0=S^1=1)
\end{equation}

Using notation from Section~\ref{sec:notation}, we estimate this quantity as:
\begin{equation}
    \hat\tau^\text{ATT} = \frac{1}{\sum_i \mathbb{I}\{A_i=1\}}\sum_{i:A_i=1}\bigg(Y_i - \hat\mu^0(\vec{x}_i)\bigg)
\end{equation}
where $\hat\mu^0(\vec{x}_i)$ is the predicted mean potential outcome under control for unit $i$, and $\mathbb{I}\{A_i=a\}$ is an indicator variable taking the value 1 when $A_i=a$ and 0 otherwise. As shown earlier, the outcome under control is set-identified given the conditional distribution of $Y$. Using our parametric outcome model, we estimate the bounds by first simulating $R$ draws from the conditional outcome distribution for each unit $i$,
\begin{equation}
    \tilde{Y}_{ir} \stackrel{\text{iid}}{\sim} \mathcal{N}\left( \hat\mu(0,\vec{x}_i), \hat\sigma^2(0,\vec{x}_i)\right), \quad r = 1,\ldots,R
\end{equation}

The nonparametric bounds retain specific quantiles of the conditional distribution. Our simulation-based estimators take the same quantities from the simulated distribution.
\begin{align}
    \hat\mu^0_\text{Lower}(\vec{x}_i) &= \frac{1}{R\pi_{S^1=1\mid S=1, A = 0}(\vec{x}_i)}\sum_{r=1}^R \tilde{Y}_{ir} \mathbb{I}\left\{ \tilde{Y}_{ir} < \text{Quantile}(\tilde{Y}_{i}, \pi_{S^1=1\mid S=1, A = 0}(\vec{x}_i)) \right\} \\
    \hat\mu^0_\text{Upper}(\vec{x}_i) &= \frac{1}{R\pi_{S^1=1\mid S=1, A = 0}(\vec{x}_i)}\sum_{r=1}^R \tilde{Y}_{ir} \mathbb{I}\left\{\tilde{Y}_{ir} > \text{Quantile}(\tilde{Y}_{i}, 1 - \pi_{S^1=1\mid S=1, A = 0}(\vec{x}_i))\right\} 
\end{align}

Finally, we construct overall bound estimates by averaging the conditional bounds across the sample.
\begin{align}
    \hat\tau^\text{ATT}_\text{Lower} &= \frac{1}{\sum_i \mathbb{I}\{A_i=1\}}\sum_{i:A_i=1}\bigg(Y_i - \hat\mu^0_\text{Upper}(\vec{x}_i)\bigg) \\
    \hat\tau^\text{ATT}_\text{Upper} &= \frac{1}{\sum_i \mathbb{I}\{A_i=1\}}\sum_{i:A_i=1}\bigg(Y_i - \hat\mu^0_\text{Lower}(\vec{x}_i)\bigg)
\end{align}

While formulas vary by estimand and assumptions, the general procedure remains the same: use the nonparametric formulas but plug in simulated conditional distributions in place of true conditional distributions and sample means in place of population means.

\subsection{Inference: Construct confidence intervals by the bootstrap}

We quantify statistical uncertainty using the nonparametric bootstrap to capture sampling variation across all model components. For each of $B$ bootstrap samples ($b = 1, \ldots, B$), we re-estimate all models and calculate bounds $\hat{\tau}_{\text{Lower},b}$ and $\hat{\tau}_{\text{Upper},b}$ (either for ATE or ATT). Our $(1-\alpha)$-level confidence interval takes the empirical quantiles of order $(\alpha/2)$ and $(1-\alpha/2)$ of the lower and upper bounds, respectively. This intersection of two one-sided confidence intervals provides coverage for the identified set---an appropriate approach for our partial identification setting where parameters of interest are ``bounded'' \citep{imbens2004confidence, zhao2019sensitivity}.

\subsection{Generalization to weighted samples}

Social surveys often employ sampling weights to account for unequal selection probabilities. The procedure above generalizes to these settings by substituting weighted sample means for unweighted sample means whenever estimating population quantities. Likewise, weights can be incorporated when estimating the parametric models for outcome existence and outcome values.

\section{Empirical illustration: The effect of parenthood on hourly wage}

We illustrate with an empirical analysis of the causal effect of parenthood on hourly wages for men and women. A long line of research has documented that becoming a parent is associated with wage losses and employment declines for women \citep{budig2001, waldfogel1997effect, staff2012explaining, gough2013review, england2016highly}. However, recent evidence suggests that the motherhood wage penalty may be disappearing over time as the effect size moves closer to zero \citep{pal2016family,buchmann2016motherhood}, though other scholars find that it is stable \citep{jee2019motherhood}. Our empirical illustration defines the wage effect among those employed regardless of motherhood and shows that evidence is consistent with a range of possible estimates---including estimates near zero and estimates far from zero---depending on the assumptions one is willing to make about who comprises this latent set of people.

We analyze data from the 1997 National Longitudinal Survey of Youth Cohort. We construct a sample of 1,985 mothers and 1,837 fathers observed in the year immediately after giving birth and a comparison group of 20,543 (25,902) person-year observations on women (men) who have not yet given birth. Our confounders include pre-parenthood characteristics: race, age, education, marital status, job tenure, work experience, and employment status and hourly wage in the previous year (Fig~\ref{fig:nlsy97}).

\begin{figure}
    \resizebox{\textwidth}{!}{\begin{tikzpicture}[x = \textwidth, y = 4in]
\node at (-.18,.5) {};
\node[anchor = east, font = {\Large\bf}] (parents) at (.3,.8) {Parents};
\node[anchor = north east, black, font = \footnotesize, align = left] at (parents.south east) {1,985 mothers\\1,837 fathers};
\node[black, font = {\Large\bf}] at (.4,.9) {Pre};
\node[black, font = {\Large\bf}] (birth) at (.6,.9) {Birth};
\node[black, font = {\Large\bf}, anchor = south] at (birth.north) {First};
\node[black, font = {\Large\bf}] at (.8,.9) {Post};
\node[black, font = \Huge] (ppre) at (.4,.8) {$\bullet$};
\node[black, font = \Huge] (pbirth) at (.6,.8) {$\bullet$};
\node[black, font = \Huge] (ppost) at (.8,.8) {$\bullet$};
\draw[->, line width = 2pt, black] (ppre) to node[midway, below, align = center] {9+\\months} (pbirth);
\draw[->, line width = 2pt, black] (pbirth) to node[midway, below, align = center] {12+\\months} (ppost);
\node[anchor = east, font = {\Large\bf}] (nonparents) at (.3,.6) {Non-Parents};
\node[anchor = north east, black, font = \footnotesize, align = left] at (nonparents.south east) {\begin{tabular}{rrr}
& person-periods & persons \\
women & 20,543 & 2,794 \\
men & 25,902 &  3,436
\end{tabular}};
\node[black, font = \Huge] (npre) at (.4,.6) {$\bullet$};
\node[black, font = \Huge] (npost) at (.8,.6) {$\bullet$};
\node[black, font = \footnotesize, anchor = north west, align = center] at (npost.north east) {allowed to\\subsequently\\have a\\child};
\draw[->, line width = 2pt, black] (npre) to node[midway, below, align = center] {21+\\months} (npost);
\node[black, anchor = north, align = center] at (.4,.35) {measure\\confounders};
\node[black, anchor = north, align = center] at (.8,.35) {measure\\outcome};
\draw[->, thick, black] (.4,.35) -- (.4,.45);
\draw[->, thick, black] (.8,.35) -- (.8,.45);
\end{tikzpicture}}
\caption{\textbf{Data structure for illustration with the NLSY97.} We begin by arranging the data so that each observation on a person takes the role of a pre-observation and is paired with the soonest post-observation on that same person at least 1.75 years into the future and no more than 6 years into the future. We label observations as non-parent observations if a first birth occurs in the window and parent observations if a birth does not occur in the window, dropping observations where the pre-period is after the first birth. We keep parent observations only if the pre- and post-observations are each no more than 3 years from the birth (28,135 childless men pairs, 2,029 father pairs, 21,704 childless women pairs, 2,098 mother pairs). All of our cases have valid reports of employment. We drop cases who reported employment but did not report a wage, so that missing wages always correspond to non-employment in our analysis, producing the analytical sample sizes reported above.}
\label{fig:nlsy97}
\end{figure}

We study two causal estimands: (1) the average effect of parenthood on employment among parents, and (2) the average effect of parenthood on wage among parents who would be employed regardless of parenthood (ATT as in Eq.~\ref{eq:tau_att}).

\subsection{Nonparametric causal assumptions and parametric statistical estimators}
\label{sec:dag}

\begin{figure}
    \centering
    \begin{tikzpicture}[x = 1in, y = .6in]
\node (x) at (-1.5, .1) {$\vec{X}$};
\node (a) at (-1.5,-1) {Motherhood};
\node (m) at (.05,-1) {Employed};
\node (y) at (1.5,-1) {Wage};
\draw[->, thick] (x) -- (a);
\draw[->, thick] (x) to[out = 0, in = 110] (y);
\draw[->, thick] (x) to[out = 0, in = 135] (m);
\draw[->, thick] (a) -- (m);
\draw[->, thick] (a) to[bend right] (y);
\draw[->, thick] (m) -- (y);
\node (u) at (.75,-.4) {$U$};
\draw[->, thick] (u) -- (y);
\draw[->, thick] (u) -- (m);
\end{tikzpicture}
    \caption{\textbf{Nonparametric causal assumptions to select confounders.} We assume that all backdoor paths between motherhood and employment and between motherhood and wage are blocked by the confounder set $\vec{X}$, which in our example includes age, education, marital status, full-time employment, job tenure, work experience, and wage and employment each lagged by one year. Our approach is valid even when $U$ exists and is unmeasured.}
    \label{fig:dag}
\end{figure}
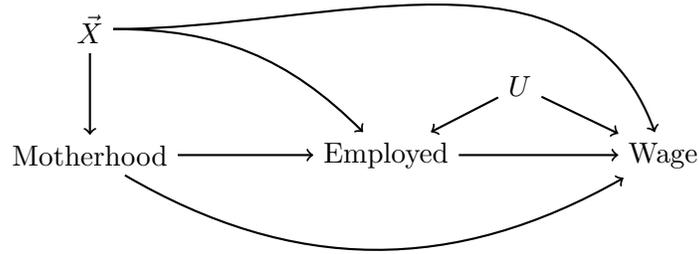

Our first causal assumption is conditional exchangeability: potential employment statuses ($S^a$) and potential wages ($Y^a$) are independent of parenthood ($A$) conditional on measured confounders $\vec{X}$. After adjusting for pre-parenthood characteristics, parenthood can be considered as-if randomly assigned with respect to potential outcomes. While untestable, this assumption is made more plausible by our rich set of pre-treatment covariates. Fig~\ref{fig:dag} presents a causal Directed Acyclic Graph (DAG) where $\vec{X}$ represents a sufficient adjustment set. The validity of this assumption relies on the richness of the available covariates and the researchers' domain knowledge about potential confounders. Importantly, our causal identification remains valid even in the presence of unobserved variables $U$ that may affect both employment and wages, such as access to advantageous job opportunities within one's social network.

We present a series of estimates under different assumptions. For women, we present estimates assuming negative monotonicity: motherhood may cause women to leave paid employment but never causes non-employed women to enter the workforce. This assumption is supported by theory, literature, and our findings (in Section~\ref{subsec:effects_on_wages}), though untestable at the individual level.

For both women and men, we consider positive mean dominance: within any covariate subgroup $\vec{x}$, potential wages are higher for always-employed individuals, on average, than for those employed under only one treatment condition. This assumption is plausible if those with stronger labor force attachment have higher wages, though one could argue that financial necessity might keep low-wage workers employed regardless of parenthood.

Our estimators follow those specified in Section \ref{sec:parametric}, using a logistic regression model for employment and a conditionally normal outcome model for log hourly wage, with quantile-based nonparametric bootstrap confidence intervals. 

\subsection{Estimated effects on employment}

\begin{figure}
    \textbf{A)} Effect of motherhood on employment\\
    \includegraphics[width=\textwidth]{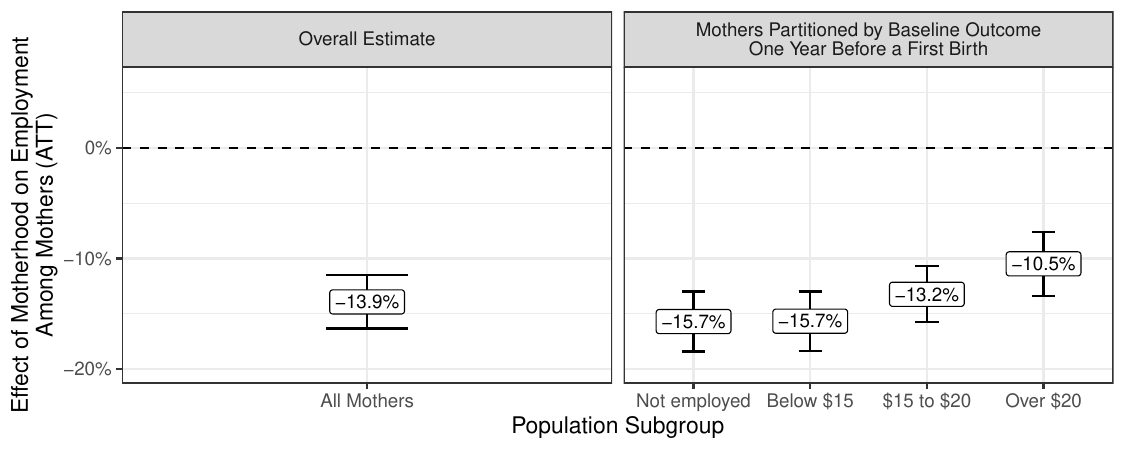} \\
    \textbf{B)} Effect of fatherhood on employment\\
    \includegraphics[width=\textwidth]{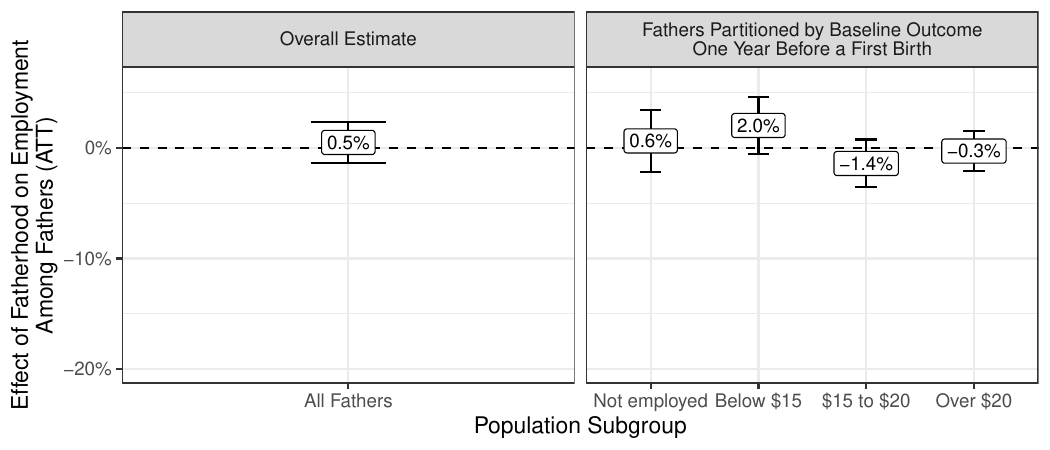}
    \caption{\textbf{Effect of parenthood on employment.} All estimates correspond to average treatment effects on the treated (effects among parents). The facet on the right groups new parents by their observed employment outcome one year before parenthood.}
    \label{fig:employment_effects}
\end{figure}

We estimate that motherhood reduces employment by 13.9 percentage points on average in our sample (Figure~\ref{fig:employment_effects}). The effect is also heterogeneous across subgroups: motherhood reduces post-birth employment most strongly among women who before birth either were not employed or were employed for less than \$15 per hour, among whom motherhood reduces employment by 15.7 percentage points. Among high-wage women whose pre-birth wages were more than \$20 per hour, motherhood reduces employment by only 10.5 percentage points. No such pattern is apparent among men, for whom fatherhood has approximately zero effect on employment in all subgroups visualized.

The pattern of employment effects supports the posited negative monotonicity assumption: motherhood either reduces employment or has no effect, but never increases it. While untestable, this assumption gains credibility from the consistently negative effect estimates in Figure~\ref{fig:employment_effects}A across population subgroups. Mean dominance assumes women whose employment is reduced by motherhood would have lower potential wages as non-mothers than those whose employment is unaffected. This assumption, though also untestable, is supported by our finding that motherhood reduces employment most among women with lower pre-birth wages.

With approximately zero employment effects for men across all subgroups (Figure~\ref{fig:employment_effects}B), monotonicity is less defensible for fathers. Given these null effects, we do not report results that assume monotonicity for fathers.

\subsection{Estimated effects on wages}\label{subsec:effects_on_wages}

\begin{figure}
    \centering
    \includegraphics[width = \textwidth]{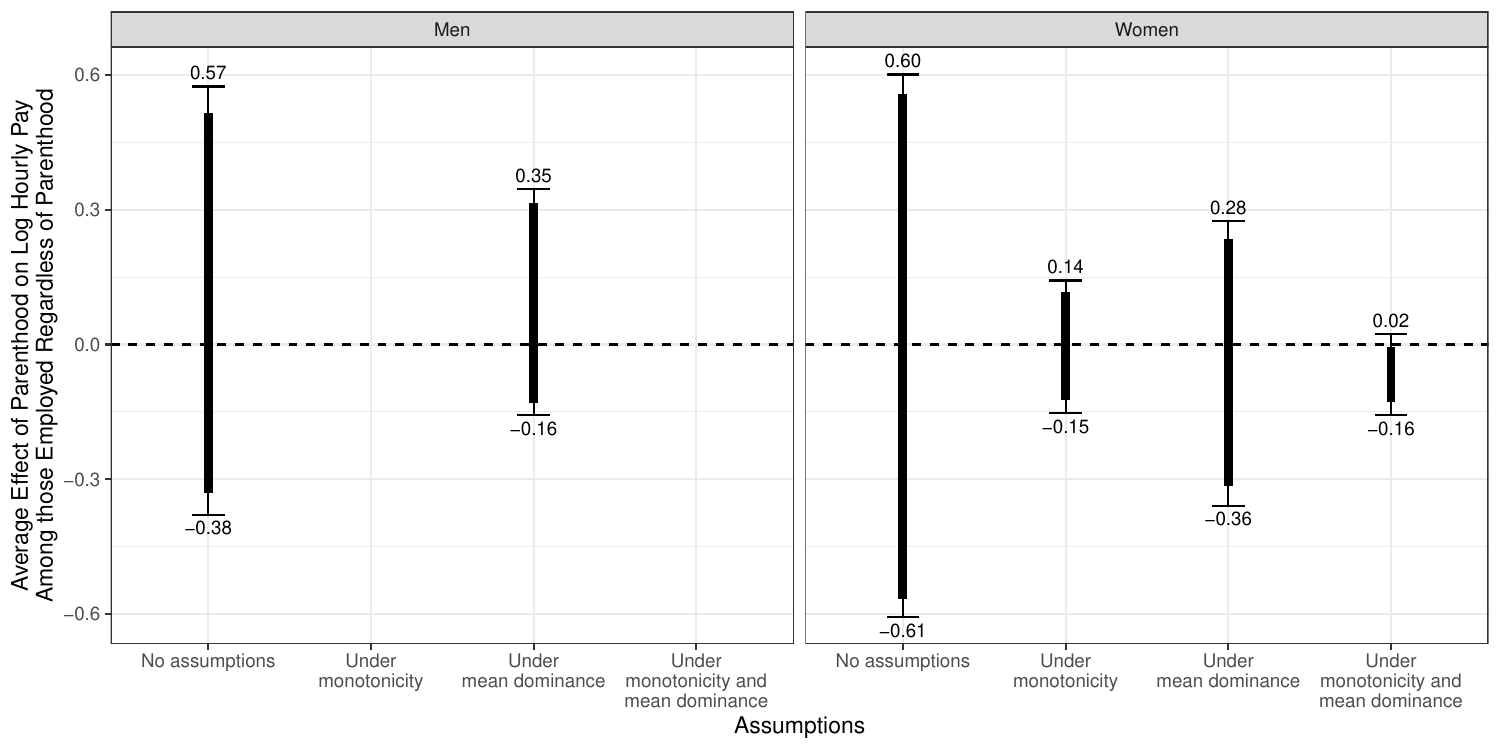}
    \caption{\textbf{Effect of parenthood on wage among those employed regardless.}}
    \label{fig:results_wage}
\end{figure}

We now illustrate bounded estimates on wages, which demonstrate that additional assumptions transparently tighten the estimated bounds \citep{manski1995identification, manski2003partial}. Figure~\ref{fig:results_wage} presents the estimated effect of parenthood on hourly wage among always-employed individuals. Under exchangeability alone, bounds are wide: fatherhood's effect on log wages ranges from -0.38 to +0.57, while motherhood's ranges from -0.61 to +0.60. These wide bounds reflect fundamental uncertainty about each person's counterfactual employment status under the treatment value that did not occur.

Adding negative monotonicity narrows bounds for women to [-0.15, 0.14], demonstrating the identifying power of theoretically-justified assumptions. For men, however, monotonicity is questionable. Some men might become employed upon fatherhood to support their family, while others (e.g., those with a high-earning spouse) might step back from employment to care for children. We therefore choose not to report estimates for men that rely on monotonicity.

Imposing only mean dominance---that always-employed men have higher potential wages---provides different estimates. Mean dominance yields bounds of [-0.16, 0.35] for fathers, suggesting a possible fatherhood premium. For mothers, this assumption alone suggests bounds of [-0.36, 0.28], moderately tighter than the baseline bounds. When combining both monotonicity and mean dominance, we obtain the tightest bounds of [-0.16, 0.02] for the motherhood wage effect. This suggests a predominantly negative effect with a wage penalty of up to 16\%.

The sequential incorporation of assumptions demonstrates both the value and limitations of theory-driven causal inference. While successive assumptions based on qualitative understanding narrow our bounds \citep{coppock2022qualitative}, substantial uncertainty (due to outcomes that may not exist) remains even under our strongest assumptions. This contrasts with standard practice: regression excluding non-employed women yields a deceptively precise estimate near zero (estimate = -0.01, CI = [-0.03, 0.02]). This false precision stems from ignoring the uncertainty about which individuals would remain employed under different treatment conditions. Our approach explicitly acknowledges this uncertainty, revealing a wider range of plausible effects. Appendix~\ref{sec:bias} formally demonstrates how analyses restricted to units with existing outcomes suffer from selection bias.

Our results contribute to the literature on parenthood wage effects by showing how selection into employment misleads inferences about wage penalties of motherhood. The bounds under various assumptions suggest standard estimates need reconsideration in light of selective employment. Moreover, the stronger employment effects among lower-wage women suggest that standard analyses may understate motherhood's total impact by focusing solely on wages conditional on employment.

\section{Discussion}

Research in social stratification regularly faces a difficult problem: treatments that shape the values of outcomes often also determine whether those outcomes exist at all. We showed how the standard practice of restricting analysis to observed outcomes obscures causal effects and produces misleading inferences about inequality. For researchers facing this problem, there are two primary recommendations. First, before studying the values of outcomes researchers should estimate and report effects of the treatment on outcome existence. When a treatment has a large effect on employment, for example, analyses that jump to wage as the outcome may miss an essential part of the story. To estimate effects on the existence of the outcome requires no new tools beyond standard causal inference methods, and could yield valuable new insights. Second, when researchers move on to study the value of selectively-existing outcomes they should explicitly focus on the latent subgroup who would have an outcome under either treatment condition. When producing results, these researchers should make transparent assumptions such as monotonicity and mean dominance, and they should report interval estimates that correctly incorporate their fundamental uncertainty about which units are in the latent stratum of interest: those whose outcome would exist under either treatment condition.

The primary contribution of this paper is showing how bounding strategies that originated in biostatistics can be adapted for use in studies of social stratification. To do so, our technical contribution is an approach involving parametric regression, simulation, and aggregation of simulated conditional estimates to marginal summaries. The parametric approach we develop enables researchers to use concepts originally designed for randomized experiments and apply them in settings where it is necessary to adjust for a large set of measured confounders.

A second contribution of this paper is to demonstrate through an empirical example that standard practice can produce misleadingly tight confidence intervals centered near zero when true effects may be much larger in magnitude. Our reexamination of the motherhood wage penalty shows that while conventional estimates produce a motherhood effect near zero with a tight confidence interval, our strongest assumptions yield bounds of [-0.16, 0.02] that suggest the possibility of a sizeable negative effect. Our bounds reveal that the motherhood wage penalty could be substantially more negative than previous research suggests---an uncertainty masked by conventional approaches that restrict analyses to employed women.

Most broadly, our framework provides a template for studying social stratification when outcome existence is itself selective. We believe there exist many processes in social stratification in which inputs that shape the values of an outcome also determine its existence. Rather than treating non-existent outcomes as missing data to be handled through deletion or imputation, their explicit incorporation can deepen understanding of inequality.

\clearpage

\section*{Preregistration statement}

The analyses reported in this paper were not preregistered.

\section*{Code, materials, and data availability statement}

Replication code for this project calls and R package that we wrote to carry out the procedures developed in this paper. The replication code and R package are currently available at an anonymized link here: \href{https://osf.io/kz24g/?view\_only=7ab62760060f4be0a9e7ed5d9e2e5f4a}{\textcolor{blue}{https://osf.io/kz24g/?view\_only=7ab62760060f4be0a9e7ed5d9e2e5f4a}}

For data access, the replication package contains a README with instructions to upload a tagset of variable to the NLSY website in order to download the raw data directly from the data distributor. The purpose of sharing data this way is so that all users of the data create their own account and agree to terms and conditions from the data distributor.

\section*{Declaration of conflicting interests}

The author(s) declared no potential conflicts of interest with respect to the research, authorship, and/or publication of this article.

\clearpage
\begin{singlespacing}
\bibliography{bib.bib}    
\end{singlespacing}

\appendix

\section{Proof of target stratum size}

\subsection{General case: No monotonicity assumption}
\label{sec:proof_no_monotonicity}

This section provides a formal proof of the set identification results for the proportion in the always-survive principal stratum ($S^0=S^1=1$) discussed in Section \ref{sec:ate_no_monotonicity}. We show that we can partially identify this quantity through bounds in the most general case without monotonicity or mean dominance. In this proof, we assume exchangeability and for simplicity assume all probabilities and expectations are conditional on measured covariates $\vec{X}$. We further define abbreviated notation for the conditional probability of membership in each principal stratum.
\begin{align}
    \pi_{11} &\equiv \P(S^0=1,S^1=1) &\text{survive regardless} \\
    \pi_{10} &\equiv \P(S^0=0,S^1=1) &\text{survive if treated} \\
    \pi_{01} &\equiv \P(S^0=1,S^1=0) &\text{survive if untreated} \\
    \pi_{00} &\equiv \P(S^0=0,S^1=0) &\text{never survive}
\end{align}

We will use two combinations of principal strata whose sizes are causally identified under conditional exchangeability.
\begin{align}
    \pi_{11} + \pi_{10} 
    &= \P(S^1=1) &\text{by definitions} \\
    &= \P(S^1=1\mid A = 1) &\text{by exchangeability} \\
    &= \P(S=1\mid A = 1) &\text{by consistency}\\
    \pi_{11} + \pi_{01} 
    &= \P(S^0=1) &\text{by definitions} \\
    &= \P(S^0=1\mid A = 0) &\text{by exchangeability} \\
    &= \P(S=1\mid A = 0) &\text{by consistency}
\end{align}

We then solve these equations to get upper bounds on $\pi_{11}$.
\begin{align}
\pi_{11} &= \P(S=1\mid A = 1) - \pi_{10}\\
&\leq \P(S=1\mid A = 1) &\text{since }\pi_{10}>=0  \\
\pi_{11} &= \P(S=1\mid A = 0) - \pi_{01}\\
&\leq \P(S=1\mid A = 0) &\text{since }\pi_{01}>=0 
\end{align}

Because these are both valid upper bounds, the upper bound is the minimum of the two. The intuition of these bounds is that the proportion who survive regardless of treatment can be no larger than the survival rate in each of the treatment conditions.
\begin{equation}
\pi_{11} \leq \text{min}\bigg(\P(S=1\mid A = 1),\P(S=1\mid A = 0)\bigg)
\label{eq:upper_limit_always}
\end{equation}

Next, we want to place a lower bound on $\pi_{11}$. For this, we begin from the fact that the probabilities of all four strata sum to 1.
\begin{align}
    1 &= \pi_{11} + \pi_{10} + \pi_{01} + \pi_{00} \\
    &= \underbrace{\pi_{11} + \pi_{10}}_\text{term 1} + \underbrace{\pi_{11} + \pi_{01}}_\text{term 2} + \pi_{00} - \pi_{11} &\text{add and subtract }\pi_{11} \\
    &= \underbrace{\P(S =1 \mid A = 1)}_\text{term 1} + \underbrace{\P(S = 1\mid A = 0)}_\text{term 2} + \pi_{00} - \pi_{11}\\
    \pi_{11} &= \P(S =1 \mid A = 1) + \P(S = 1\mid A = 0) - 1 + \pi_{00} \\
    &\geq \P(S = 1\mid A = 1) + \P(S = 1\mid A = 0) - 1 &\text{since }\pi_{00}\geq 0 \label{eq:lower_limit_always_1}
\end{align}

In the result above, two limiting cases help to build intuition. For the first limiting case, suppose the survival rate is 100\% among the treated and 100\% among the untreated, so that everyone survives regardless of treatment. The lower bound on the proportion who survive regardless is (1 + 1 - 1) = 1. For the second limiting case, suppose that the survival rates in the treated and control conditions sum to 1. For example, suppose that 75\% of the treated survive and 25\% of the untreated survive. In this case, the entire population may consist of treatment-induced-surivors (75\% of the population) and control-induced-survivors (25\% of the population). It is possible that no one is in the survive-regardless group. In math, the lower limit on the proportion always surviving is (0.75 + 0.25 - 1) = 0.

Finally, we also know $\pi_{11} \geq 0$ since it is a probability. Together with Eq.~\ref{eq:lower_limit_always_1}, this gives a lower limit on the proportion always survivors. Paired with Eq.~\ref{eq:upper_limit_always}, we have partial identification for the proportion who survive regardless of treatment.
\begin{equation}
\text{max}\bigg\{0,\P(S = 1\mid A = 1) + \P(S = 1\mid A = 0) - 1\bigg) \leq \pi_{11}\leq \text{min}\bigg(\P(S=1\mid A = 1),\P(S=1\mid A = 0)\bigg\}    
\label{eq:bound_always_1}
\end{equation}

An intuition for the lower bound would be helpful. If $\P(S = 1\mid A = 1) + \P(S = 1\mid A = 0) > 1$, then there must be some overlap in the groups that survive under treatment and control. The amount of necessary overlap is exactly $\P(S = 1\mid A = 1) + \P(S = 1\mid A = 0) - 1$. If $\P(S = 1\mid A = 1) + \P(S = 1\mid A = 0) \leq 1$, then it's possible that there's no overlap (i.e., no one survives regardless of treatment), so the lower bound is 0. For example, If 80\% survive under treatment and 70\% survive under control, the sum is 150\%, meaning at least 50\% must survive in both conditions. If 60\% survive under treatment and 40\% survive under control, the sum is 100\%, so it's possible that no one survives in both conditions.

In the data analyzed, we restrict to those who survive ($S = 1$). This implies that we need the proportion always-survivors conditional on survival.
\begin{align}
    \P(S^0=S^1=1\mid S = 1, A = a) 
    &= \frac{\P(S^0=S^1=1\mid A = a)}{\P(S = 1\mid A = a)} &\text{def. of cond. prob.} \\
    &= \frac{\P(S^0=S^1=1)}{\P(S = 1\mid A = a)} &\text{conditional exchangeability} \\
    &= \frac{\pi_{11}}{\P(S = 1\mid A = a)}
    \label{eq:s11_given_survival}
\end{align}
Because the denominator can be estimated from data, the bounds on the numerator imply bounds on the fraction.

\subsection{Under Monotonicity}
We now consider the target stratum size under negative monotonicity $(S^0 \geq S^1)$, which in our motherhood example means that motherhood may cause a woman to leave employment but never cause employment. Under negative monotonicity, it holds that $\pi_{10}=0$. With that, we can point-identify $\pi_{11}$:
\begin{align}
\pi_{11} + \pi_{10} &= \P(S=1\mid A = 1) \\
\Rightarrow \pi_{11} &= \P(S=1\mid A = 1) &\text{since }\pi_{10} = 0
\end{align}

Now let's determine the proportion of always-survivors conditional on survival for both treatment groups, based on Eq.~\ref{eq:s11_given_survival}:
\begin{align}
\P(S^0=S^1=1|S=1,A=a) =
 \begin{cases}
 1 & \text{if } a = 1 \\
 \frac{\P(S=1|A=1)}{\P(S=1|A=0)} & \text{if } a = 0
 \end{cases}
\end{align}

This makes intuitive sense under negative monotonicity: when $a=1$, anyone who survives despite treatment would certainly have survived without it. When $a=0$, only a fraction of those who survive without treatment would survive with treatment---this fraction equals the ratio of the survival rates.

Similarly, we can easily get the same quantity under positive monotonicity case:
\begin{align}
\P(S^0=S^1=1|S=1,A=a) =
\begin{cases}
\frac{\P(S=1|A=0)}{\P(S=1|A=1)} & \text{if } a = 1 \\
1 & \text{if } a = 0
\end{cases}
\end{align}

\section{Derivation of bias in naive comparison under Monotonicity}\label{sec:bias}

To formally see why naive comparisons of wages between employed mothers and non-mothers can be misleading, we derive the bias term explicitly. For simplicity, we focus on the case under monotonicity (that motherhood never increases employment) and suppress conditioning on observed covariates $\vec{X}$ in our notation, though this conditioning would be necessary in practice for identification in observational setting. We assume monotonicity in the direction that motherhood does not increase employment: $S^0 \geq S^1$, meaning a woman who would be employed as a mother would also be employed as a non-mother.

Let $\hat{\tau}_{naive}$ be the difference in mean wages between employed mothers and employed non-mothers. Note that under monotonicity, employed non-mothers represent a mixture of the always-employed stratum and those employed only when not mothers, hence $\P(S^0=1) = \P(S^0=S^1=1) + \P(S^0=1,S^1=0)$.
\begin{align}
\hat{\tau}_{naive} &= \E(Y \mid S=1, A=1) - \E(Y \mid S=1, A=0) \\
&= \E(Y^1 \mid S^1=1) - \E(Y^0 \mid S^0=1) &\text{by consistency} \\
&= \E(Y^1 \mid S^0=S^1=1) - \E(Y^0 \mid S^0=1) &\text{by monotonicity} \\
&= \E(Y^1 \mid S^0=S^1=1) \nonumber \\ 
& \qquad - \bigg( \E(Y^0 \mid S^0=S^1=1) \frac{\P(S^0=S^1=1)}{\P(S^0=1)} \nonumber \\
& \qquad + \E(Y^0 \mid S^0=1,S^1=0)\frac{\P(S^0=1,S^1=0)}{\P(S^0=1)} \bigg),
\end{align}
where the last equality is by conditioning on $S^1$ for the second term. The causal estimand, the average causal effect among the employed-regardless, we seek to identify is: $$\tau=\E(Y^1-Y^0 \mid S^0=S^1=1).$$
Thus, the bias in the naive comparison can be written as:
\begin{align}
Bias(\hat{\tau}_{naive}, \tau) &= \E( \hat{\tau}_{naive} - \tau )\\
&= \underbrace{[\E(Y^0 \mid S^0=S^1=1) - \E(Y^0 \mid S^0=1,S^1=0)]}_{(1)} \underbrace{\frac{\P(S^0=1,S^1=0)}{\P(S^0=1)}}_{(2)}
\end{align}

This bias term shows why the naive comparison may not recover the causal effect even under monotonicity. The bias is the product of two quantities: (1) the difference in potential wages under non-motherhood between the always-employed and those who would be employed only as non-mothers, and (2) the proportion of employed non-mothers who would leave upon motherhood. The sign of the bias term depends on whether women who maintain employment through motherhood have higher or lower potential wages, on average, compared to those who would leave employment upon motherhood.

This derivation also speaks to the implication of mean dominance. Mean dominance would assume that women who maintain employment through motherhood have higher potential wages even under non-motherhood compared to those who would exit employment upon becoming mothers. Under mean dominance, our bias term becomes surely positive since $E(Y^0 \mid S^0=S^1=1) - E(Y^0 \mid S^0=1,S^1=0) \geq 0$, suggesting that naive comparisons would understate the size of the motherhood wage penalty.

\end{document}